\documentclass[11pt]{article}
\newcommand{\gcite}[2]{\hbox{\cite{#1}--\cite{#2}}}

\begin{document}
\LARGE
\begin{center}
\bf{An interpolation model for the equation of state in binary systems.}
\end{center}
\Large
   \begin{center}
   S.A.Kukushkin{\,}\footnote{e-mail: ksa@math.ipme.ru },
   A.V.Osipov{\,}\footnote{e-mail: oav@math.ipme.ru},
   Yu.V.Pavlov,
   A.S.Sokolov \\
   \end{center}
   \begin{center}
\large {\em Institute of Mechanical Engineering, Russian Academy of Sciences, \\
   61 Bolshoy, V.O., St.Petersburg, 199178, Russia }
   \end{center}
\hrule
\begin{abstract}
The new interpolation model of state of binary mixture is investigated. This
model use only two parameters and produce many type of phase diagrams.
\end{abstract}
\hrule
\vspace{9mm}

As it is known, at present the virial equation of state is the only
theoretically substantiated form for an equation of gaseous
state~\cite{1}.
   However, application of this equation to description of behavior
of real gases, especially within the range of high densities, presents certain
difficulties, since it involves a great number of terms, to be taken into
account for virial expansion.
   Therefore, in practice, various empirical equations of state are often
   used.
In elementary case the equations include two parameters, which can be
estimated by the known values of critical magnitudes for temperature,
pressure and volume.
  The Van der Waals equation (see, for instance~\cite{2})
and the Redlich-Kwong equation~\cite{3} are the most applicable as the
two-parametric equations.
    The Redlich-Kwong model was further extended in papers of
G.M.Wilson, G.Soave and others \gcite{4}{7}, where the authors proposed to add a
certain function, the so-called $\alpha$-function, dependent on the
characteristic temperature.
   One of the main limitations in such modified models is the presence
of a great number of parameters, which cannot be rather correctly proved.
   In the case of analysis of a gaseous mixture, the number of parameters
is to be increased.
    Consequently, to solve this problem, it is necessary to construct
a model containing a minimum number of parameters to be defined, and , at
the same time, it would be possible to describe the behavior of a real
gas' mixture up to a liquid state within the whole range of changes
in pressure and temperature by means of the model.
    Within the framework of the model presented below,  the equation of
state for a binary gas mixture is treated, where together with two
the Van der Waals parameters of components the only additional parameter
of interaction is introduced.
    Meanwhile, as it is shown below, even within the framework of a such
simplified statement the basic types of phase equilibria can be examined
and most of phase diagrams presently known can be obtained.

    As it is known, the Van der Waals free energy for a gas is defined
by the formula (see, for instance, \cite{8}):
    \begin{equation}
F = N \, T \, \ln \frac{N}{V-N b} - \frac{N^2 a}{V} + N \, f(T)
\label{1}
\end{equation}
    In this work we put forward the following formula of free energy for
a mixture of two gases, namely,
    \begin{eqnarray}
F = T \, \left[N_1 \, \ln \frac{N_1}{V-(N_1 b_1 - N_2 b_2)} + N_2 \, \ln
\frac{N_2}{V-(N_1 b_1 - N_2 b_2)}\right] - \nonumber\\
- \left[\frac{N^2_1 \, a_1}{V} +
\frac{N^2_2 \, a_2}{V}\right] + \tilde{A} \frac{N_1 N_2}{V} + (N_1 +N_2)
\, f(T)       \phantom{xxxxx}
\label{2}
\end{eqnarray}
    Here $N_i$ and $b_i$ are the matter content and molecular "volume"
of the $i$-th component, respectively;
$a_i$ is the positive constant characterizing the interaction of
the $i$-th gaseous component.
   The free energy defined by (\ref{2})  can be represented in the form of
$$
F = F_1 + F_2 + F_{int} \, ,
$$
    where
$F_{int} = \tilde {A} \, N_1 N_2 / V$ , whereas $F_i$ is
the free energy of the $i$-th gaseous component in the presence of
another component.

    Further, we confine ourselves to analysis of a symmetric model,
i.e. we assume that
$a_1 = a_2 = a$   and   $b_1 = b_2 = b$.
   In addition, in analysis of phase equilibria the last terms
in (\ref{2}) can be neglected, since it does not contribute to the
equations for pressure and chemical potentials of different phases.

    Now we introduce the dimensionless values, according to formulas
$$
V_r = V / V_{crit} \ ,  \ \  T_r = T / T_{crit} \ , \ \ \rho_i
= N_i / V_r \ ,
$$
where the critical values of volume and temperature are determined by
the following expressions:
$V_{crit} = 3 b$,\ $T_{crit} = 8 a/(27 b)$.\
Then, the expression for the reduced density of free energy
$\tilde{F} = F (81 b^2) / (8 a V) $
takes the form
    \begin{eqnarray}
\tilde{F} ({\rho}_1, {\rho}_2 \, T) &=& T \, \left[{\rho}_1 \,
\ln \frac{{\rho}_1}{1 - ({\rho}_1 + {\rho}_2) /3} + {\rho}_2
\, \ln \frac{{\rho}_1}{1 - ({\rho}_1 + {\rho}_2) /3}\right]
-  \nonumber\\ &-& \frac{9}{8} \, [{\rho}^2_1 + {\rho}^2_2] +
\tilde{A} {\rho}_1 {\rho}_2  \,.
\label{3}
\end{eqnarray}

    When instead of gaseous component densities ${\rho}_i$
the total density $\rho$ and ''concentration`` $x$ of the first
component by formulae
$\rho = {\rho}_1 + {\rho}_1$, \, $x= ({\rho}_1 -
{\rho}_2) /({\rho}_1 + {\rho}_2)$, $x \in [-1,1]$
are introduced, the expression for the density of free energy in
the above new variables takes the form:
   \begin{eqnarray}
\tilde{F} (T,\rho,x) &=& T \, \rho \, \ln \frac{\rho}{1 -
\rho / 3 } - \frac{9}{8} \, {\rho}^2 + \alpha {\rho}^2
(1-x^2) +    \\
&+& \frac{T \rho}{2} \, [(1+x) \ln (1+x) + (1-x) \ln (1-x)]  \,.
\label{4}  \nonumber
\end{eqnarray}
    In the presented expression the parameter
$\alpha = \tilde{A}/4 + 9/16$ is responsible for the interaction energy of
gaseous components. It is easily seen that in the case with an one-component gas
($x \to \pm 1$) eqn. (\ref{4})  takes the form of the Van der Waals equation
(1).

     From the presented expression for the free energy the formulae for
the chemical potentials of gaseous components and pressure are
directly followed, namely:
\begin{equation}
{\mu}_1 = T \, \ln
\left(\frac{\rho (1+x)}{1 - \rho /3}\right)
+ \frac{T}{1 - \rho /3}
- \frac{9}{4} \, \rho + 2 \alpha \rho (1-x)    \,,
\label{5}
\end{equation}
    \begin{equation}
{\mu}_2 (\rho, x) = {\mu}_1 (\rho, -x)  \,,
\label{6}
\end{equation}
    \begin{equation}
P = \frac{T \rho}{1 - \rho /3} - \frac{9}{8} \, {\rho}^2
+ \alpha {\rho}^2 (1-x^2)    \,.
\label{7}
\end{equation}

    By means of (5) -- (7) one can construct phase diagrams with
different values of parameter $\alpha$.
   Let the pressure in a system be specified.
   The equality of chemical potentials for gaseous components
$$
\mu_1(\rho, x, T)=\mu_2(\rho, x, T)
$$
is in agreement with conditions of ''gas-gas`` or ''liquid-liquid``
phase equilibria at the specified pressure $P(\rho,x,T) = P_0$.

    The explicit form of the set of equations for the ''gas-gas`` or
''liquid-liquid`` equilibria is
    \begin{equation}
\frac{T \rho}{1 - \rho /3} - \frac{9}{8} \, {\rho}^2 +
\alpha {\rho}^2 (1-x^2) = P_0 \ ,\ \ \
T \, \ln \frac{1+x}{1-x} = 4
\alpha \rho x    \,.
\label{8}
\end{equation}

    The equality of equilibria in the both phases,
$P ({\rho}_g,x_g,T) = P ({\rho}_l,x_l,T) = P_0$
and the equality of chemical potentials of the binary system
components are the conditions for ''gas-liquid`` phase equilibrium, namely,
$$
{\mu}_1 ({\rho}_g,x_g,T) = {\mu}_1 ({\rho}_l,x_l,T)  \ , \ \ \
{\mu}_2 ({\rho}_g,x_g,T) = {\mu}_2 ({\rho}_l,x_l,T)  \,,
$$
where lower indices  $g$ and $l$ refer to gaseous and liquid phases,
respectively.

\begin{center}
{\bf Discussion}
\end{center}

     The set of equations (8) and corresponding system for the
''gas-liquid`` equilibrium are investigated both numerically and
analytically.
    With different values of determinative
parameters of the problem $\alpha$ and $P_0$, various types of diagrams can be
obtained, such as, the so-called cigar, pinched cigar, the diagram of eutectic
type and so on.
    In the case with large values of parameters $\alpha$ and $P_0$,
the curves of phase equilibrium are of domal shape and they correspond to
the ''liquid-liquid`` phase equilibrium.
   Here, at $x \to \pm 1$ that conforms to the case with an one-component
substance, $F_{int} \to 0$, and the ''liquid-liquid`` phase equilibrium
is observed at $T \to 0$.
   As  $x \to 0$  (with large values $\alpha$),
the contribution to the free energy of an exchange component grows fast and the
phase equilibrium is observed at higher temperature.

    When the positive interaction energy $F_{int} \to 0$
the phase diagrams are represented in the form of eutectic diagrams.
    This is, first of all, associated with the behavior of free energy
curves for gas and liquid phases during changes in the interaction energy (by
the analogy with weak solutions).

    In the case with negative values of $\alpha$  there are no ''gas-gas``
or ''liquid-liquid`` equilibrium states, that is directly followed from
mathematical analysis of the set of equations (8), i.e.
at $\alpha \le 0$, solution of the second equation in the system is
absent.

    So, by means of the present model for a mixture of nonideal gases the
authors succeeded in obtaining various types of phase equilibrium diagrams
occurred during an experiment.

{\bf Acknowledgments}. This work was supported in part by the Russian
Foundation for Basic Research (projects No 98-03-32791 and 99-03-32768),
and the Russian Federal Center ''Integration``
(project No A0151).

\end{document}